**Title**: 3D Augmented Reality-Assisted CT-Guided Interventions: System Design and Preclinical Trial on an Abdominal Phantom using HoloLens 2

**Authors**: Brian J. Park, MD, MS (1), Stephen J. Hunt, MD, PhD (1), Gregory J. Nadolski, MD (1), Terence P. Gade MD, PhD (1)

1. Perelman School of Medicine at the University of Pennsylvania, 3400 Spruce St, Philadelphia, PA 19104

**Summary Statement**: Holographic 3D AR guidance can result in safer and more efficient CT-guided interventions that require less radiation.

**Key Results**:

- Total needle passes and radiation dose significantly reduced using augmented reality
- Complication rate of hitting a nontargeted lesion are abrogated using augmented reality
- Augmented reality elevated the performance of all operators to the same level irrespective of prior clinical experience

**Abbreviations**: AR = augmented reality, DLP = dose-length product, IR = interventional radiology, 3D = three-dimensional

## Abstract

**Background:** Out-of-plane lesions pose challenges for CT-guided interventions. Augmented reality (AR) headset devices have evolved and are readily capable to provide virtual 3D guidance to improve CT-guided targeting.

**Purpose:** To describe the design of a three-dimensional (3D) AR-assisted navigation system using HoloLens 2 and evaluate its performance through CT-guided simulations.

**Materials and Methods**: A prospective trial was performed assessing CT-guided needle targeting on an abdominal phantom with and without AR guidance. A total of 8 operators with varying clinical experience were enrolled and performed a total of 86 needle passes. Procedure efficiency, radiation dose, and complication rates were compared with and without AR guidance. Vector analysis of the first needle pass was also performed.

**Results:** Average total number of needle passes to reach the target reduced from 7.4 passes without AR to 3.4 passes with AR (54.2% decrease, p=0.011). Average dose-length product (DLP) decreased from 538 mGy-cm without AR to 318 mGy-cm with AR (41.0% decrease, p=0.009). Complication rate of hitting a non-targeted lesion decreased from 11.9% without AR (7/59 needle passes) to 0% with AR (0/27 needle passes). First needle passes were more nearly aligned with the ideal target trajectory with AR versus without AR (4.6° vs 8.0° offset, respectively, p=0.018). Medical students, residents, and attendings all performed at the same level with AR guidance.

**Conclusions:** 3D AR guidance can provide significant improvements in procedural efficiency and radiation dose savings for targeting challenging, out-of-plane lesions. AR guidance elevated the performance of all operators to the same level irrespective of prior clinical experience.

## Introduction

Augmented reality (AR) technologies are able to seamlessly merge virtual objects with the surrounding environment. Extensive technological progress has been made with AR headset devices since 2006 with the development of one of the first AR guidance systems, called RAMP, for CT-guided interventions via custom headset with video overlay [1]. Since then, many other applications have been developed with AR devices to enhance training and image-guided procedures [2, 3].

However, despite technical achievements with headsets now capable of holographic overlay with see-through views of the real world, clinical utility and adoption of AR technologies have made only marginal progress since RAMP. Still, the goals of any navigation system for improving CT-guided interventions remain the same. Challenging lesions and anatomy can make targeting difficult and result in prolonged procedure times, increased radiation exposure, and more complications [4, 5]. Out-of-plane approaches, in particular, often present with challenging characteristics [6].

Many AR-assisted guidance systems have been recently developed for percutaneous needle-based interventions with the widespread availability of commercial AR devices. Smartphone or tablet-based AR navigation platforms for CT-guided needle insertion have shown sub-5 mm accuracies, decreased procedure times, and fewer intermediate CT scans [7, 8]. 3D AR-assisted navigation systems using HoloLens (v1, Microsoft, Redmond, WA) have received unanimously positive feedback among operators for its potential to enhance safety, aid in execution, and improve depth perception and spatial understanding [9, 10]. Although such systems show promise, no prior study to date has demonstrated the procedural effects of AR guidance using a headset device through a systematic trial.

This study describes the design of a 3D AR-assisted navigation system using the next-generation HoloLens 2 (Microsoft, Redmond, WA) headset device. Unlike some other existing AR-assisted navigation systems, no additional or extra hardware components are needed aside from the headset. Registration was performed automatically to a CT grid routinely used in clinical practice, as opposed to using separate external image-based markers or matrix barcodes required in other systems. Evaluation was performed through a preclinical trial simulating CT-guided needle targeting of a challenging lesion on an abdominal phantom with and without AR guidance. Procedure efficiency, radiation dose, and complication rates are compared.

## Methods

This study is Institutional Review Board exempt as no actual patient data is obtained or analyzed. CT-guided percutaneous needle targeting was simulated on a phantom model (071B, CIRS, Norfolk, VA) containing multiple targets of various sizes. A CT grid (Guidelines 117, Beekley Medical, Bristol, CT) commonly used in clinical practice was placed on the anterior surface of the phantom for planning and to serve as a fiducial target for registration.

<u>Preoperative Imaging and 3D Modeling</u>

A preoperative CT scan of the phantom was performed with 120 kVp and 2 mm slice thickness on Siemens SOMATOM Force (Fig. 1). An 11 mm lesion was selected for targeting. Manual and semi-automated segmentations of the lesions, CT grid and bony structures, and skin surface were performed with ITK-SNAP using threshold masking and iterative region growing [11]. Segmentation meshes were exported in STL file format followed by mesh decimation using Meshmixer (Autodesk, San Rafael, CA) to eliminate redundant vertices and reduce mesh size to improve 3D rendering performance. Reduced meshes were then exported in OBJ file format and material textures, including colors and transparencies, were applied using Blender (Amsterdam, Netherlands). The target lesion was colored in green; all other nontargeted lesions were colored in red. The final 3D surface-rendered model was exported in FBX file format (Fig. 2).

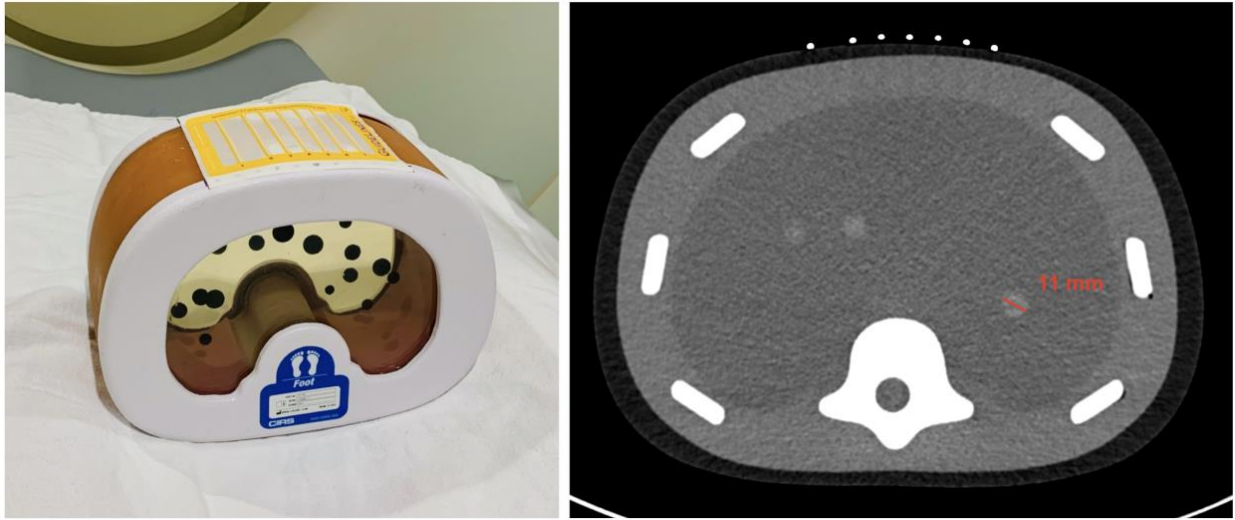

**Figure 1.** CT phantom abdominal biopsy phantom. *Left* - CT grid is applied to the surface of the model. Phantom contains multiple targets of various sizes. *Right* - CT image of model. Selected target measures 11 mm in diameter.

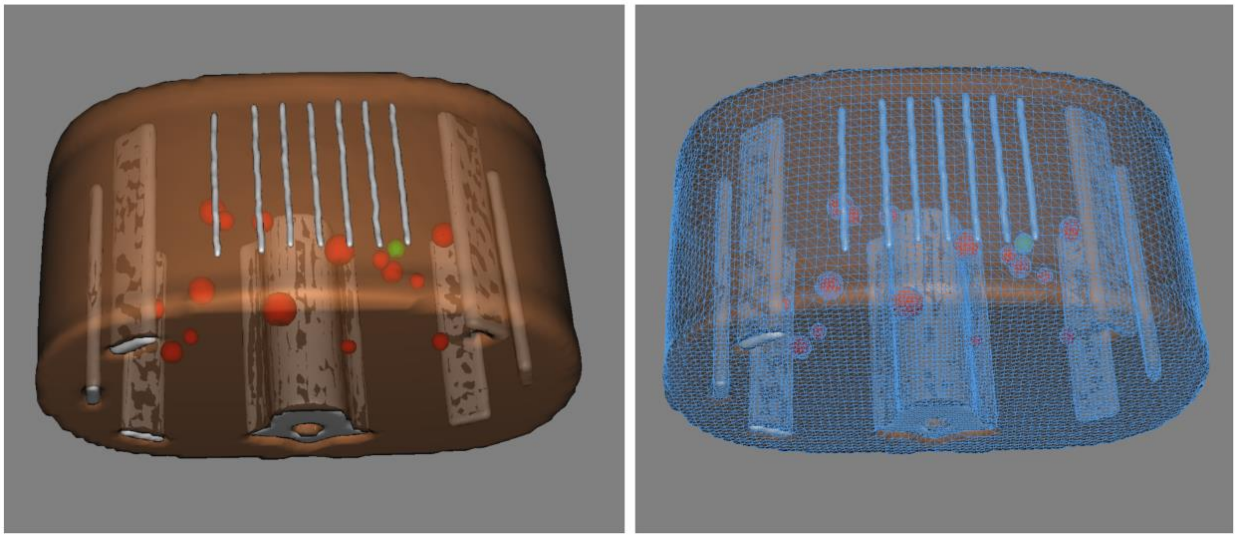

**Figure 2.** Three-dimensional surface-rendered model of phantom. *Left* - Lines from the CT grid can be seen along the anterior surface. Target lesion is specified in green. All other nontargeted lesions are specified in red. *Right* - Wireframe view of model which contains 58,498 polygons with a total file size of only 1.6 MB.

Target Trajectory

A long, out-of-plane trajectory with a narrow-window access was intentionally chosen to the 11-mm target from a skin entry site along the inferior aspect between CT gridlines 3 and 4 (Fig. 3). This trajectory angle was beyond the maximum gantry tilt for potential compensation by the CT scanner.

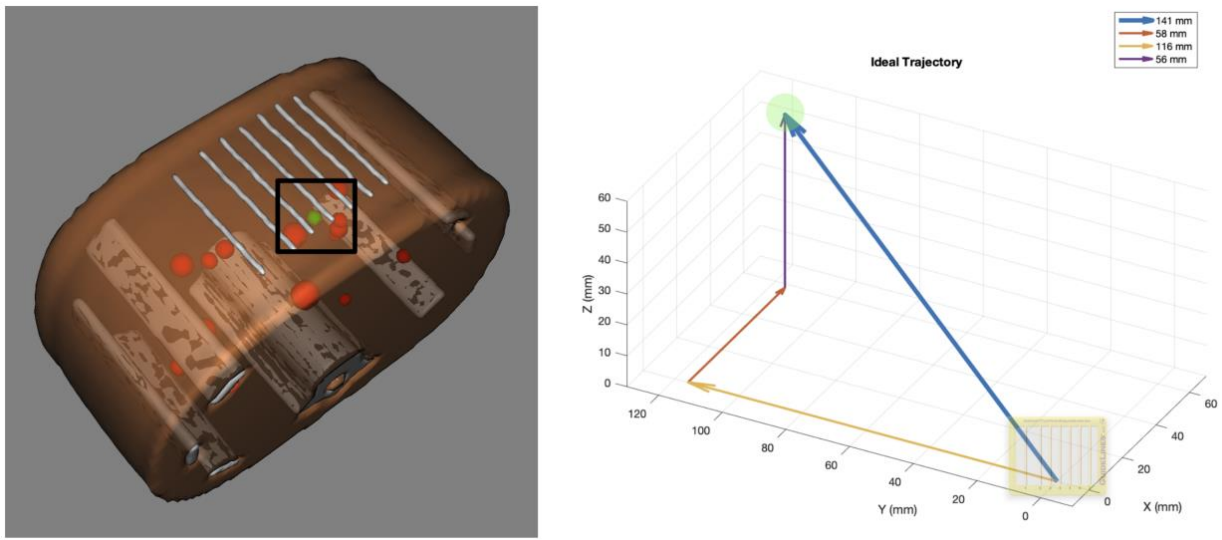

**Figure 3.** Trajectory to targeted lesion from specified skin entry site. *Left* - Down-the-barrel look at trajectory to targeted lesion (green) from skin entry site at the inferior aspect between labeled gridlines 3 and 4 (black box). Several nontargeted lesions (red) can be seen in close proximity to the trajectory. *Right* - Vector of ideal trajectory based on preoperative CT scan from specified skin entry site. Total trajectory distance of 14.1 cm from skin with 23.4° angle relative to the z-plane (5.8 cm lateral, 11.6 cm deep, and 5.6 cm cranial component). Target and CT grid are not drawn to scale.

Augmented Reality System

3D AR visualization and interaction were performed using HoloLens 2 headset device. A custom HoloLens application was developed in Unity 2019.2.21 and Mixed Reality Toolkit Foundation 2.3.0. Automated registration of the 3D model to CT grid was performed using computer vision and Vuforia 9.0.12 with the CT grid as the image target. Features on the CT grid can be reliably and quickly detected by Vuforia [12], and studies have validated the accuracy of Vuforia on HoloLens (v1) [2]. This method of fast and accurate registration is fully automated and does not require any user input, which is ideal for inexperienced HoloLens 2 users. A virtual needle trajectory was added into the 3D model based on the ideal trajectory. This virtual guide allowed the user to easily trace the ideal trajectory using a real needle (Fig. 4).

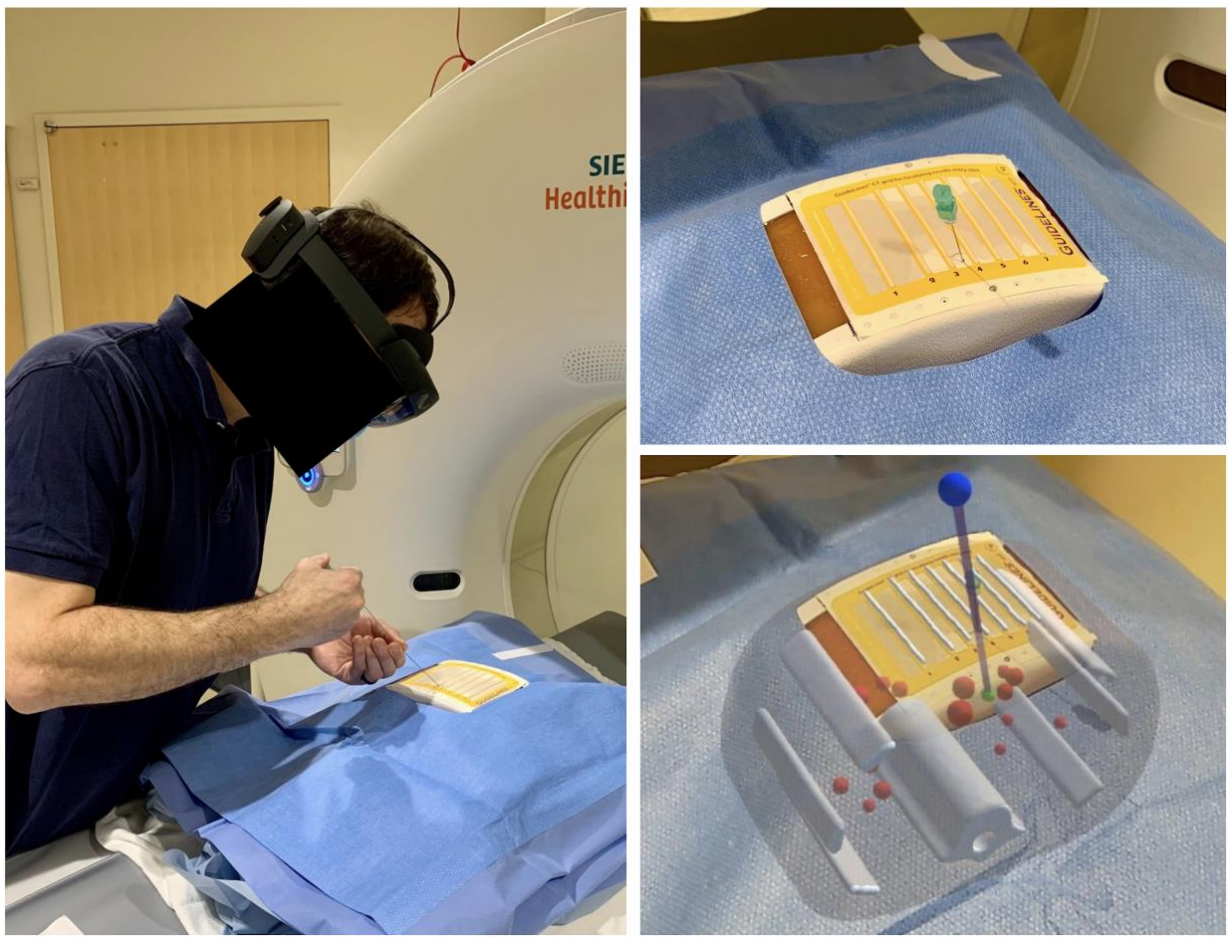

**Figure 4.** Augmented reality (AR)-assisted navigation using HoloLens 2. *Left* - Participant inserts the needle while wearing HoloLens 2. *Top right* - View of needle insertion without AR. *Bottom right* - View of needle insertion through HoloLens 2 with three-dimensional model and virtual needle guide projected onto the phantom. Registration appears accurate with the actual CT gridlines aligned with the virtual

gridlines. The needle is seen aligned with the virtual guide (purple line) displaying the ideal trajectory to the target lesion (green ball). Note that this two-dimensional captured image does not fully represent the three-dimensional stereoscopic view from HoloLens 2.

### CT-Guided Procedure Simulation

All simulations were performed on a Siemens SOMATOM Force CT scanner at 120 kVp and 2 mm slice thickness. After applying a surgical drape over the phantom, percutaneous CT-guided targeting using a 21G-20cm Chiba needle was simulated in the same fashion that is clinically performed at [omitted]. Following a topogram, an initial CT scan of the phantom was performed and reviewed for trajectory planning. The needle was then passed into the phantom and iteratively advanced, redirected, or retracted, as needed, until the tip of the needle was in the target. Interval CT scans were performed following any needle adjustment. Each adjustment was counted as a needle pass, and these passes were cumulatively documented.

A total of 8 participants simulated CT-guided needle targeting: 2 attendings, 3 interventional radiology (IR) residents, and 3 medical students. Both attendings had greater than 5 years of experience. 2 residents were in their final year of training. All 3 medical students had never previously seen nor performed a CT-guided intervention. Aside from 1 resident, all other participants had no prior experience wearing or interacting with HoloLens 2. In order to limit bias, participants were randomized into cohorts: CT-guided targeting 1) without AR and then repeated with AR or 2) with AR and then repeated without AR (Fig. 5).

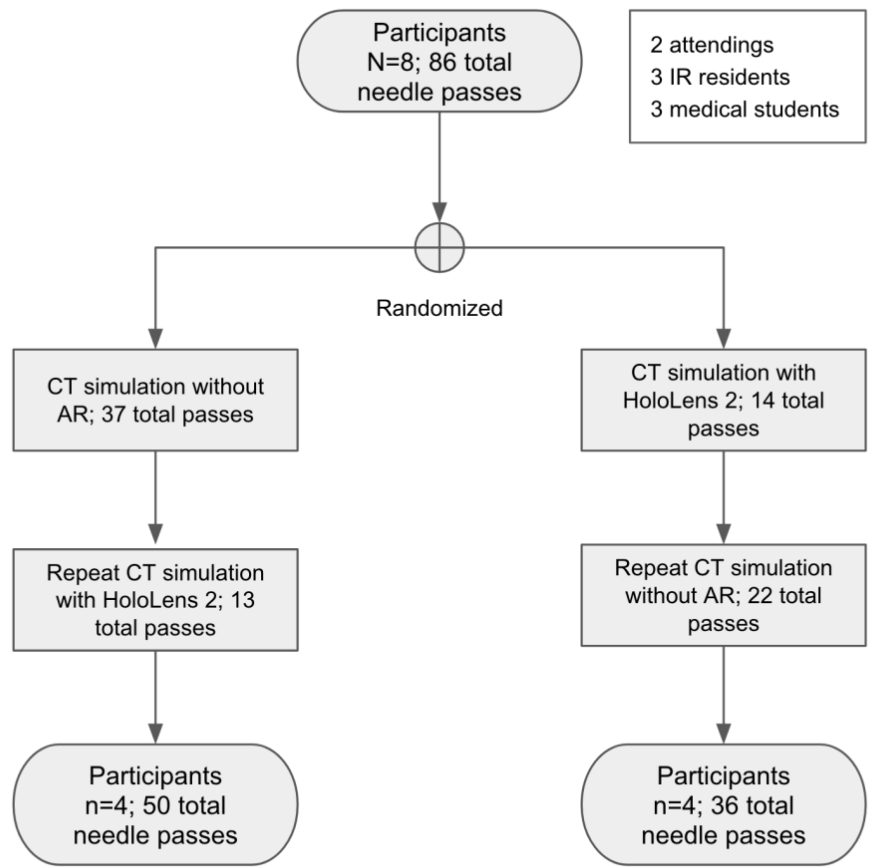

**Figure 5.** Flowchart of study design. Order of interventions with and without augmented reality-assisted navigation were randomized to limit order bias.

Procedural Imaging and Vector Analysis

Total number of needle passes were recorded. Total DLP was obtained from the CT dose report. Vector analysis of the CT scan after the first, initial needle pass was performed (Fig. 6). These CT scans were resampled into isotropic volumes (1x1x1 mm) using 3D Slicer 4.10.1 and linear interpolation [13]. Voxel locations at the skin entry site, needle tip, and target centroid were recorded. Distances and angles were calculated using vector magnitude and dot product, respectively. All CT scans were reviewed to determine any complications. A complication was defined as any needle pass that unintentionally hits or goes through a nontargeted lesion.

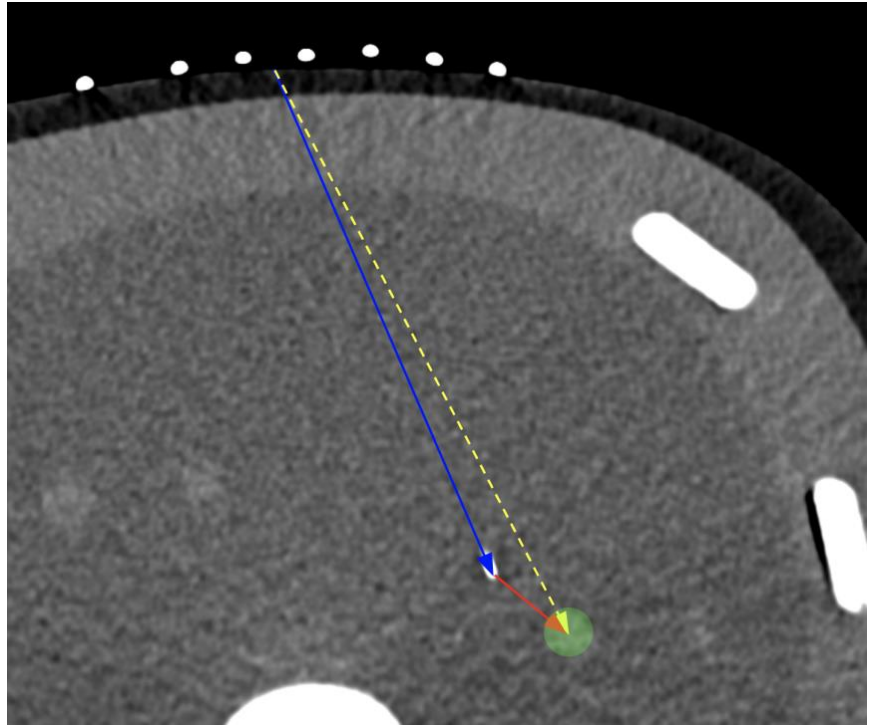

**Figure 6.** Diagram demonstrating calculations in two dimensions for illustrative purposes only. Actual calculations were performed in three dimensions based on voxel locations. Blue solid arrow represents distance of needle tip from skin entry site. Red solid arrow represents remaining distance to center of target. Yellow dotted arrow represents ideal trajectory from skin entry site to center of target. Angle offsets were calculated between the needle trajectory (blue solid arrow) relative to the ideal trajectory (yellow dotted arrow).

Statistical Analysis

Vector analysis, averages, and paired t-tests were performed using Google Sheets (Mountain View, CA). Post hoc power analysis suggested a total sample size of 8 for a power of 0.8 and effect size of 1 to achieve a statistical significance level of 0.05.

## Results

A comparison of CT-guided needle targeting performed with and without AR is summarized in Table 1. The use of AR-assisted guidance significantly reduced the number of iterative passes required to reach the selected target, from an average of 7.4 passes without AR down to 3.4 passes with AR (54.2% decrease, p=0.011). Radiation dose savings were also significant with average DLP decreasing from 538 mGy-cm without AR to 318 mGy-cm with AR (41.0% decrease, p=0.009). The complication rate of hitting a nonselected target also decreased from 11.9% (7/59 passes) without AR to 0% with AR (0/27 passes).

The first, initial needle pass angle was more aligned with the ideal trajectory with AR guidance compared to needle insertion without using any guidance as is common clinical practice (4.6° vs 8.0° respectively, p=0.018). In addition, the 1st pass distance traveled was deeper and the needle tip was closer to the target with AR versus without AR but did not reach statistical significance.

**Table 1. AR-assisted CT-guided Percutaneous Needle Targeting (N=8; 86 passes)**

| Procedural metric | Without AR (59) | With AR (27) | % change | p-value |
|---|---|---|---|---|
| Avg # of passes until selected target reached | 7.4 | 3.4 | 54.2 | 0.011* |
| Avg DLP (mGy-cm) | 538 | 318 | 41.0 | 0.009** |
| Complication rate of hitting nonselected target | 11.9% (7) | 0% (0) | | |
| Avg 1st pass angle offset from ideal trajectory (degrees) | 8.0 | 4.6 | 42.8 | 0.018* |
| Avg 1st pass distance from skin (cm) | 9.6 | 10.2 | 5.9 | 0.763 |
| Avg 1st pass distance remaining to target (cm) | 5.4 | 4.0 | 26.5 | 0.330 |

*p<0.05, **p<0.01

Subgroup analysis was performed comparing results based on prior clinical experience (Table 2). Without AR guidance, the complication rate decreased with increasing clinical experience; medical students had the highest total number of passes that hit a nonselected target, and attendings had zero passes that hit a nonselected target. Medical students also had the greatest 1st pass traveled distance of 14.7 cm compared to residents and attendings with 6.4 cm and 6.7 cm, respectively.

Using 3D AR guidance, all subgroups performed similarly. The average number of passes ranged between 3-4 passes, and total 1st pass distance averaged approximately 10 cm among each subgroup. Additionally, there were no complications of hitting a nonselected target with AR. Medical students with no prior clinical experience performed at the same level as experienced attendings with AR.

| Table 2. Subgroup Analysis by Clinical Experience | | | | | | |
|---|---|---|---|---|---|---|
| Experience | Without AR | | | With AR | | |
| | Avg # of passes | Total complications | Avg 1st pass distance from skin (cm) | Avg # of passes | Total complications | Avg 1st pass distance from skin (cm) |
| Attendings (2) | 9.0 | 0 | 6.7 | 3.5 | 0 | 10.4 |
| IR Residents (3) | 8.3 | 2 | 6.4 | 3.3 | 0 | 10.1 |
| Medical Students (3) | 5.3 | 5 | 14.7 | 3.3 | 0 | 10.0 |

## Discussion

3D AR-assisted guidance showed significant improvements in efficiency and radiation dose savings for CT-guided targeting of a challenging, out-of-plane lesion. AR guidance decreased the total number of needle passes by 54.2%. Fewer needle passes also resulted in fewer interval CT scans, leading to a 41.0% decrease in total radiation dose. This decrease in radiation does not quite match the decrease in needle passes since an initial CT scan was performed for planning at the start of all simulations. Furthermore, with AR guidance, the 1st needle pass was more in line with the ideal trajectory compared with no guidance. This likely contributed to the fewer number of subsequent, iterative needle adjustments required to reach the target.

Some degree of order bias and recall did occur during non-AR-assisted simulations. The cohort that performed targeting with HoloLens 2 first and then repeated without AR had fewer total needle passes without AR (22 passes) compared to the cohort that performed targeting without AR first (37 passes). These data suggest that visually seeing the trajectory in 3D in addition to physically performing the procedure concurrently may enhance spatial understanding and innate recall [14]. However, these effects were not realized during AR-assisted simulations, which overall had consistent findings regardless of the order of interventions between the cohorts (13 vs 14 passes). This suggests that the benefits of having real-time 3D navigation likely supersede advantages associated with prior experience or recall. This contention is further supported by the fact that medical students performed at the same level as experienced attendings with 3D AR guidance.

Generally, medical students were aggressive on their first needle pass without AR, advancing the needle a distance of over twice that of residents and attendings with a goal of getting close to the target as opposed to multiple smaller passes to ensure the trajectory is on course, which may come with clinical experience. As expected, medical students had the most complications of hitting nontargeted lesions. Without AR, residents and attendings took more conservative 1st initial passes and required a greater number of overall passes but had fewer complications, with attendings having no complications. With AR guidance, attendings, residents, and students all performed at the same level with similar total passes and 1st-pass distances as well as having zero complications.

The primary limitation of this study was the evaluation of this system using a stationary, inanimate phantom. As with any navigation system, patient motion, respiratory breathing, soft tissue deformation, and needle bending are important factors that can affect navigational performance. Breathing can be compensated by techniques such as simple respiratory gating [15] or high frequency jet ventilation under general anesthesia [16]. Soft tissue deformation can be compensated by deformable modeling [10], and needle bending can be extrapolated using a shape sensing needle [17].

In summary, 3D AR guidance using a headset device can improve efficiency, reduce radiation dose, and minimize complications during CT-guided interventions. The use of 3D AR guidance may thus facilitate the treatment of challenging, hard-to-reach, or out-of-plane lesions. Additionally, these data suggest that AR guidance may immediately help to elevate the performance of inexperienced operators, providing added opportunities to treat challenging lesions that were previously declined due to limited operator experience. Although this preclinical trial shows promising and translatable benefits of 3D AR guidance, further developments and robust clinical testing will be needed for adoption into actual practice.